# On Providing Downlink Services in Collocated Spectrum-Sharing Macro and Femto Networks[1]

Xiaoli Chu[2], Yuhua Wu, David López-Pérez, and Xiaofeng Tao

*Abstract* – Femtocells have been considered by the wireless industry as a cost-effective solution not only to improve indoor service providing, but also to unload traffic from already overburdened macro networks. Due to spectrum availability and network infrastructure considerations, a macro network may have to share spectrum with overlaid femtocells. In spectrum-sharing macro and femto networks, inter-cell interference caused by different transmission powers of macrocell base stations (MBS) and femtocell access points (FAP), in conjunction with potentially densely deployed femtocells, may create dead spots where reliable services cannot be guaranteed to either macro or femto users. In this paper, based on a thorough analysis of downlink (DL) outage probabilities (OP) of collocated spectrum-sharing orthogonal frequency division multiple access (OFDMA) based macro and femto networks, we devise a decentralized strategy for an FAP to self-regulate its transmission power level and usage of radio resources depending on its distance from the closest MBS. Simulation results show that the derived closed-form lower bounds of DL OPs are tight, and the proposed decentralized femtocell self-regulation strategy is able to guarantee reliable DL services in targeted macro and femto service areas while providing superior spatial reuse, for even a large number of spectrum-sharing femtocells deployed per cell site.

*Index Terms* – Downlink, femtocell, macrocell, OFDMA, outage probability, spectrum-sharing.

## I. INTRODUCTION

Almost all current cellular networks are facing problems arising from imperfect service providing, especially indoors. One cost-effective solution to improve service providing and network capacity is the emerging femto network, where low-power miniature base stations (BSs), a.k.a. femtocell access points (FAPs) [1], home BSs [2], or home eNodeBs [3], are overlaid on macro networks. Each FAP provides high-data-rate wireless connections to user equipments (UEs) within a short range using the same radio-access technology as the macro underlay. FAPs are connected to an operator's network via local broadband connections. Due to concerns about security, backhaul capacity and customer preference [4], femtocells are likely to be deployed in a closed-access mode [2], [5], i.e., a femtocell serves only a group of authorized UEs. Orthogonal frequency division multiple access (OFDMA) based femtocells are widely anticipated to deliver massive improvements in coverage and capacity for next generation mobile networks [1], [6].

---

[1] This research was supported by the UK EPSRC Grants EP/H020268/1, CASE/CNA/07/106, and EP/G042713/1. Part of this work was presented at *IEEE VTC Fall'11*, San Francisco, USA, September 2011.
[2] Corresponding author: x.chu@sheffield.ac.uk

Inter-cell interference is among the most urgent challenges to successful rollouts of femtocells [2]. A centralized downlink (DL) frequency planning across OFDMA-based femto and macro cells was proposed in [7]. However, as plug-and-play devices, the number and locations of active FAPs would be hardly known to operators, so interference in femto networks cannot be managed by centralized network planning [6],[8]. In [9], cross-tier DL interference is avoided by assigning orthogonal spectra to the macro and femto tiers, and femto-to-femto interference is mitigated by allowing each femtocell to access only a random subset of the frequency sub-channels assigned to the femto tier. Although operating femtocells in a dedicated spectrum can eliminate the interference to and from the macro network, operators may still choose to deploy both macro and femto networks in a common spectrum due to considerations of spectrum availability, cost, and network infrastructure [7], [10], [11]. In spectrum-sharing macro and femto networks, different transmit powers used by macro BSs (MBSs) and FAPs, in conjunction with potentially densely deployed closed-access femtocells, may create dead spots where reliable DL services cannot be guaranteed to either macro UEs (MUEs) or femto UEs (FUEs). For example, the DL of a femtocell that is close to an MBS may be disrupted by macro DL transmissions due to the much higher transmit power used by the MBS; while the DL of an MUE that is far away from its serving MBS may be blocked by DL transmissions of nearby femtocells [11]. A carrier-sensing based interference mitigation strategy, where an FAP monitors macro uplink pilot transmissions and adjusts transmit power based on its distance from the MBS upon sensing an MUE in its vicinity, was proposed for spectrum-sharing macro and femto networks in [12].

Probability-distribution analysis of macro and femto DL signal-to-interference ratios (SIRs) plays an essential role in [9], [12], [13], where it is assumed that all FAPs transmit at a same fixed power level. However, it is anticipated that FAPs will need to dynamically adjust their transmit power levels in the radio resources that they share with other cells, so as to mitigate inter-cell interference. Moreover, the analytical DL SIR distributions in [9] account for macro-to-macro or femto-to-femto interference only, and are not in closed-form expressions. The same target DL SIR and outage probability (OP) constraint are used for a macrocell and a femtocell in [12], whereas higher data rates are typically required by indoor UEs than outdoor UEs. The effect of shadowing is not considered in [12], while the effect of fading is not considered in [13].

In this paper, we investigate DL service providing problems in spectrum-sharing macrocell and closed-access femtocells, starting with an analysis of their DL OPs, based on which we derive analytical expressions of the minimum distance that a femtocell has to keep away from an MBS, and distance-dependent upper and lower bounds on FAP transmit power for maintaining reliable macro and femto DL services. We then propose a decentralized strategy for FAPs to self-regulate their transmit power and usage of radio resources depending on their distances from the closest MBS, so as to ensure satisfactory macro and femto DL services over targeted regions.

What distinguish this work from existing results is summarized as follows. First, our analysis of macro and femto DL OPs removes the assumption that all FAPs transmit at the same power, embraces the dynamics of transmit powers used by different FAPs instead, and allows different SIR targets and OP constraints for macro and femto cells. Second, our DL OP analysis accounts for path loss, Rayleigh fading and lognormal (LN) shadowing, and decomposes the femto DL OP into two parts corresponding to femto DL outages caused by strong macro-to-femto interference alone and those caused by composite macro-and-femto interference. Third, macro and femto DL OPs are obtained in closed-form lower bounds, thereby permitting explicit studies of the impact of system and channel parameters on co-channel deployment of macro and femto networks. Last, the decentralized femto self-regulation strategy requires only infrequent updates of an FAP's distance from the closest MBS, MBS transmit power, channel statistics, and a local spatial density of femtocells, which can be provided by the operator through the backhaul.

Simulation results show that our closed-form macro and femto DL OP lower bounds are tight, and the decentralized femto self-regulation strategy is able to guarantee satisfactory macro and femto DL services in their respectively targeted service areas while providing much improved spatial reuse. In the rest of the paper, system and channel models are introduced in Section II, DL OP analysis is presented in Section III, the femto self-regulation strategy is proposed in Section IV, simulation results are provided in Section V, and conclusions are drawn in Section VI.

## II. SYSTEM AND CHANNEL MODELS

We consider the OFDMA DL of collocated spectrum-sharing macro and femto networks. The basic radio resource unit that can be allocated in OFDMA transmissions is a resource block (RB) [3]. Intra-cell interference is avoided by scheduling at most one UE per RB in each cell [9], [14]. We assume that all subcarriers of an RB are assigned with the same power. The macrocell of interest serves outdoor UEs in a disc area centered at the MBS with a radius $r_M$. We assume that the macrocell is fully loaded with all its available RBs in use at any time. Interference from its neighboring macrocells is ignored for analytical tractability [12], [15]. Closed-access indoor femtocells, each serving a number of authorized indoor UEs over a disc area centered at the FAP with a radius $r_F$, are randomly overlaid on the macrocell. FAPs' locations at a point in time are modeled by a homogeneous spatial Poisson point process (SPPP) $\Omega$ with an intensity of $\lambda_F$ on the $R^2$ plane [16], [17], which provides a tractable spatial-distribution model for suburban or rural residential femtocells. A single-antenna transceiver is assumed for each MBS, FAP, and UE.

Because the bandwidth and time duration of an RB are restricted [3], it is assumed that all subcarriers within an RB experience the same LN shadowing and frequency-flat Rayleigh fading [9], the shadowing and fading coefficients remain constant within each RB [18], but may vary from one RB to the next. As cellular networks are interference limited by nature, thermal noise at receivers is neglected for analytical simplicity. We use the IMT-2000 channel model [19] for terrestrial radio propagation decays. Path losses for links from an MBS to an outdoor UE, from a serving indoor FAP to its indoor UE, from an indoor FAP to an outdoor UE, from an MBS to an indoor UE, and from an interfering indoor FAP to an indoor UE are given respectively in Table I.

III. DOWNLINK OUTAGE PROBABILITIES

*A. Femtocell Downlink Outage Probability*

Under the assumption that all subcarriers of an RB are assigned with the same power and see the same channel state, the SIR of an RB is equivalent to that of one of its subcarriers. For an indoor FUE located at the cell edge of its serving FAP, the received SIR of an RB is given by

$$\text{SIR}_F = \frac{P_F \phi_F^{-1} r_F^{-\alpha_F} H_F Q_F}{P_M \phi_{FM}^{-1} D_{FM}^{-\alpha_{FM}} H_{FM} Q_{FM} + \sum_{i \in \Phi} P_{Fi} \phi_{FF}^{-1} D_{FFi}^{-\alpha_{FF}} H_{FFi} Q_{FFi}} \qquad (1)$$

where interference is summed over transmissions from the MBS and the set $\Phi$ of FAPs that are transmitting in the considered RB; since the set of FAPs transmitting in the given RB excluding the FAP of interest has the same spatial distribution as the set $\Phi$ [16], we use $\Phi$ hereafter to simplify notation; $P_F = P_{F,Tx} G_F G_U$, being $P_{F,Tx}$ the transmit power per subcarrier of the serving FAP, $G_F$ the FAP antenna gain, and $G_U$ the UE antenna gain; $P_M = P_{M,Tx} G_M G_U$, being $P_{M,Tx}$ the MBS transmit power per subcarrier and $G_M$ the MBS antenna gain; $P_{Fi} = P_{Fi,Tx} G_F G_U$ ($i \in \Phi$), being $P_{Fi,Tx}$ the transmit power per subcarrier of interfering FAP $i$; $G_F$, $G_U$ and $G_M$ are assumed the same for all FAPs, UEs and MBSs, respectively; $\phi_F$, $\phi_{FM}$ and $\phi_{FF}$ are fixed radio propagation losses, and $\alpha_F$, $\alpha_{FM}$ and $\alpha_{FF}$ are path-loss exponents, as defined in Table I; the femtocell radius $r_F$ is used for the worst case scenario; $D_{FM}$ and $D_{FFi}$ are distances from the MBS and interfering FAP $i$ to the FUE, respectively; $H_F$, $H_{FM}$ and $H_{FFi}$ denote the exponentially distributed unit-mean channel power gains [9] respectively from the serving FAP, MBS and interfering FAP $i$ to the FUE; $Q_F \sim \text{LN}(\zeta \mu_F, \zeta^2 \sigma_F^2)$, $Q_{FM} \sim \text{LN}(\zeta \mu_{FM}, \zeta^2 \sigma_{FM}^2)$ and $Q_{FFi} \sim \text{LN}(\zeta \mu_{FF}, \zeta^2 \sigma_{FF}^2)$ are the LN shadowing from the serving FAP, MBS and interfering FAP $i$ to the FUE, respectively, with $\zeta = 0.1\ln10$ [9] and means ($\mu_F$, $\mu_{FM}$, $\mu_{FF}$) and standard deviations ($\sigma_F$, $\sigma_{FM}$, $\sigma_{FF}$) are all in decibels (dB). The random variables (RVs) $H_F$, $Q_F$, $H_{FM}$, $Q_{FM}$, $H_{FFi}$, $Q_{FFi}$, $D_{FM}$ and $D_{FFi}$ are independent.

Dynamic transmit power levels of FAPs are typically upper-limited by a maximum allowable level and would also be lower-limited for making transmissions detectable. From the perspective of a victim UE, it is reasonable to model the unknown transmit power of an interfering FAP as an RV with finite mean and variance. However, it is not a trivial task to find the distribution that statistically characterizes dynamic transmit power of FAPs, which would require measurement campaigns taking into account various factors such as femtocell deployment scenario, radio resource management, network loading, etc. In [20], it was reported that output power (in mW) of most contemporary wireless communication systems approximately exhibits LN distribution. The same observation was also made in [21] and was later verified in [22], [23]. Based on these results, we use LN distribution to model the unknown transmit power (in mW) of an interfering FAP. For analytical tractability, we further assume that the probability distributions of transmit power levels per subcarrier from different interfering FAPs are independent and approximately identical, i.e., $P_{Fi,\text{Tx}} \sim \text{LN}(\zeta\mu_{F,\text{Tx}}, \zeta^2\sigma_{F,\text{Tx}}^2)$, $\forall i \in \Phi$, where $\mu_{F,\text{Tx}}$ and $\sigma_{F,\text{Tx}}$ are in dBm. Accordingly, $(P_{Fi,\text{Tx}})_{\text{dBm}}$ is approximately normally distributed. If $P_{F,\text{Tx,min}}$ and $P_{F,\text{Tx,max}}$ are the minimum and maximum power levels that an FAP transmits in a subcarrier, then following the *empirical rule* [24], we have $(P_{F,\text{Tx,min}})_{\text{dBm}} \approx \mu_{F,\text{Tx}} - 3\sigma_{F,\text{Tx}}$ and $(P_{F,\text{Tx,max}})_{\text{dBm}} \approx \mu_{F,\text{Tx}} + 3\sigma_{F,\text{Tx}}$.

As a basic quality-of-service (QoS) requirement for an indoor FUE, its instantaneous SIR has to be no less than an SIR target $\gamma_F$. Denoting $S_F = P_F\phi_F^{-1}r_F^{-\alpha_F}H_FQ_F$ the received signal power from the serving FAP and $I_{FM} = P_M\phi_{FM}^{-1}D_{FM}^{-\alpha_{FM}}H_{FM}Q_{FM}$ the received interference power from the MBS, and assuming identical channel statistics across all RBs, the DL OP of an indoor FUE is given by

$$\Pr(\text{SIR}_F < \gamma_F) = \Pr\left(\frac{S_F}{I_{FM} + \sum_{i\in\Phi}P_{Fi}\phi_{FF}^{-1}D_{FFi}^{-\alpha_{FF}}H_{FFi}Q_{FFi}} < \gamma_F\right) = \Pr\left(\frac{S_F}{I_{FM}} < \gamma_F\right) + \Pr\left(\text{SIR}_F < \gamma_F, \frac{S_F}{I_{FM}} \geq \gamma_F\right) \quad (2)$$

where $\Pr(S_F/I_{FM} < \gamma_F)$ is the probability of macro-to-femto interference being strong enough to cause an FUE outage, and $\Pr(\text{SIR}_F < \gamma_F, S_F/I_{FM} \geq \gamma_F)$ is the probability of femto-to-femto interference together with not-strong-enough macro-to-femto interference causing an FUE outage. Thus, the analysis of femto DL OP can be decomposed into two sub-problems, which facilitate gaining insights into how the femto DL OP is affected by macro and/or femto interference, respectively.

For an FUE at a distance $d$ from the MBS, the first probability in the last line of (2) is given by

$$\Pr\left(\frac{S_F}{I_{FM}} < \gamma_F \Big| D_{FM} = d\right) = \Pr\left(\frac{P_F\phi_F^{-1}r_F^{-\alpha_F}H_FQ_F}{P_M\phi_{FM}^{-1}D_{FM}^{-\alpha_{FM}}H_{FM}Q_{FM}} < \gamma_F \Big| D_{FM} = d\right)$$
$$= \Pr\left(\frac{H_FQ_F}{H_{FM}Q_{FM}} < \frac{P_M\phi_F r_F^{\alpha_F}\gamma_F}{P_F\phi_{FM}d^{\alpha_{FM}}}\right) \quad (3)$$

Since the product of an exponential RV and a LN RV can be approximated by another LN RV [9],[25], we approximately have $H_F Q_F \sim \text{LN}(\tilde{\mu}_F, \tilde{\sigma}_F^2)$, with $\tilde{\mu}_F = \zeta(\mu_F - 2.5)$ dB, $\tilde{\sigma}_F = \zeta\sqrt{\sigma_F^2 + 5.57^2}$ dB, and $H_{FM} Q_{FM} \sim \text{LN}(\tilde{\mu}_{FM}, \tilde{\sigma}_{FM}^2)$, with $\tilde{\mu}_{FM} = \zeta(\mu_{FM} - 2.5)$ dB, $\tilde{\sigma}_{FM} = \zeta\sqrt{\sigma_{FM}^2 + 5.57^2}$ dB. Letting $\vartheta = H_F Q_F / (H_{FM} Q_{FM})$, it is easy to show that $\vartheta \sim \text{LN}(\tilde{\mu}_F - \tilde{\mu}_{FM}, \tilde{\sigma}_F^2 + \tilde{\sigma}_{FM}^2)$, and (3) can be calculated by using the LN cumulative distribution function (CDF) of $\vartheta$ as follows

$$\Pr\left(\frac{S_F}{I_{FM}} < \gamma_F \bigg| D_{FM} = d\right) \approx F_\vartheta\left(\frac{P_{M,Tx} G_M \phi_F r_F^{\alpha_F} \gamma_F}{P_{F,Tx} G_F \phi_{FM} d^{\alpha_{FM}}}; \tilde{\mu}_F - \tilde{\mu}_{FM}, \sqrt{\tilde{\sigma}_F^2 + \tilde{\sigma}_{FM}^2}\right) \quad (4)$$

where $F_\vartheta(x; \tilde{\mu}_F - \tilde{\mu}_{FM}, \sqrt{\tilde{\sigma}_F^2 + \tilde{\sigma}_{FM}^2}) = \Pr(\vartheta \leq x)$ is the CDF of the LN RV $\vartheta$.

The second probability in the last line of (2) can be written as

$$\Pr\left(\text{SIR}_F < \gamma_F, \frac{S_F}{I_{FM}} \geq \gamma_F\right) = \int_0^\infty \int_0^\infty \Pr\left(\frac{w}{z + \sum_{i \in \Phi} P_{Fi} \phi_{FF}^{-1} D_{FFi}^{-\alpha_{FF}} H_{FFi} Q_{FFi}} < \gamma_F, \frac{w}{z} \geq \gamma_F\right) dF_{S_F}(w) dF_{I_{FM}}(z) \quad (5)$$

where $F_{S_F}(w) = \Pr(S_F \leq w)$ and $F_{I_{FM}}(z) = \Pr(I_{FM} \leq z)$ are the CDFs of $S_F$ and $I_{FM}$, respectively.

For any received femto signal power $w\ (> 0)$ and macro interference power $z\ (> 0)$, and under the condition of $w/z \geq \gamma_F$ (i.e., the macro interference alone will not cause an outage at the FUE), we can split the set $\Phi$ of FAPs into two disjoint complementary sets, i.e., $\Phi = \Phi_{w,z} \cup \Phi_{w,z}^c$, with

$$\Phi_{w,z} = \left\{i \in \Phi \bigg| \frac{w}{z + P_{Fi} \phi_{FF}^{-1} D_{FFi}^{-\alpha_{FF}} H_{FFi} Q_{FFi}} < \gamma_F, \frac{w}{z} \geq \gamma_F\right\} \text{ and } \Phi_{w,z}^c = \left\{i \in \Phi \bigg| \frac{w}{z + P_{Fi} \phi_{FF}^{-1} D_{FFi}^{-\alpha_{FF}} H_{FFi} Q_{FFi}} \geq \gamma_F, \frac{w}{z} \geq \gamma_F\right\},$$

where $\Phi_{w,z}$ is the set of dominant interfering FAPs to the indoor FUE. Hence, a lower bound of (5) is given by the probability that at least one dominant interfering FAP, together with the MBS, contributes enough interference to individually cause a femto DL outage relative to $\gamma_F$, i.e.,

$$\Pr\left(\text{SIR}_F < \gamma_F, \frac{S_F}{I_{FM}} \geq \gamma_F\right) \geq \int_0^\infty \int_0^\infty \Pr(\Phi_{w,z} \neq \varnothing) dF_{S_F}(w) dF_{I_{FM}}(z) \quad (6)$$

where $\Pr(\Phi_{w,z} \neq \varnothing)$ is the probability of the marked SPPP set $\Phi_{w,z}$ being nonempty [17]. A lower bound obtained in this way is generally tight even if the spatial intensity $\lambda_F$ of the homogeneous SPPP $\Omega$ is large, because it is unlikely that a large group of interferers could collaboratively cause an outage without at least one of them being a dominant interferer [17].

As $H_F Q_F \sim \text{LN}(\tilde{\mu}_F, \tilde{\sigma}_F^2)$ and $H_{FM} Q_{FM} \sim \text{LN}(\tilde{\mu}_{FM}, \tilde{\sigma}_{FM}^2)$, it is easy to show that $S_F \sim \text{LN}(\tilde{\mu}_F, \tilde{\sigma}_F^2)$ approximately, with $\breve{\mu}_F = \tilde{\mu}_F + \ln(P_F \phi_F^{-1} r_F^{-\alpha_F})$, and $I_{FM}(d) = P_M \phi_{FM}^{-1} d^{-\alpha_{FM}} H_{FM} Q_{FM}$ follows also a LN distribution, i.e., $I_{FM}(d) \sim LN(\breve{\mu}_{FM}, \tilde{\sigma}_{FM}^2)$, with $\breve{\mu}_{FM} = \tilde{\mu}_{FM} + \ln(P_M \phi_{FM}^{-1} d^{-\alpha_{FM}})$. Substituting the LN probability density functions (PDFs) of $S_F$ and $I_{FM}(d)$ into (6), we calculate $\Pr(SIR_F < \gamma_F, S_F/I_{FM} \geq \gamma_F | D_{FM} = d)$ for an indoor FUE located at a distance $d$ from the MBS and obtain

$$\Pr\left(SIR_F < \gamma_F, \frac{S_F}{I_{FM}} \geq \gamma_F \middle| D_{FM} = d\right) \geq \sum_{n=1}^{N} \sum_{m=1}^{M} \frac{w_n v_m \left\{1 - \exp\left[-\kappa_F \lambda_F \left(e^{\breve{\mu}_F + \sqrt{2a_n + 2\bar{\chi}(b_m)}\tilde{\sigma}_F} - \gamma_F e^{\sqrt{2}\tilde{\sigma}_{FM} b_m + \breve{\mu}_{FM}}\right)^{-\frac{2}{\alpha_{FF}}}\right]\right\}}{2\pi \sqrt{a_n + \bar{\chi}(b_m)} e^{\bar{\chi}(b_m)}} \quad (7)$$

for which the full derivation is provided in Appendix A, under the assumption that each FAP transmits in an RB with a 100% probability, where $N$, $w_n$ and $a_n$ ($n = 1, ..., N$) are the order, weight factors and abscissas of the Laguerre integration [26], $M$, $v_m$ and $b_m$ ($m = 1, ..., M$) are the order, weight factors and abscissas of the Gauss-Hermite integration [26], respectively,

$$\kappa_F = \pi \left(\frac{G_F G_U \gamma_F}{\phi_{FF}}\right)^{\frac{2}{\alpha_{FF}}} \exp\left[\frac{2(\tilde{\mu}_{FF} + \zeta \mu_{F,Tx})}{\alpha_{FF}} + \frac{2(\tilde{\sigma}_{FF}^2 + \zeta^2 \sigma_{F,Tx}^2)}{\alpha_{FF}^2}\right], \text{ and } \bar{\chi}(y) = \frac{\left(\sqrt{2}\tilde{\sigma}_{FM} y + \breve{\mu}_{FM} + \ln \gamma_F - \breve{\mu}_F\right)^2}{2\tilde{\sigma}_F^2},$$

and $H_{FFi} Q_{FFi} \sim LN(\tilde{\mu}_{FF}, \tilde{\sigma}_{FF}^2)$ ($\forall i \in \Phi$), with $\tilde{\mu}_{FF} = \zeta(\mu_{FF} - 2.5)$ dB, $\tilde{\sigma}_{FF} = \zeta \sqrt{\sigma_{FF}^2 + 5.57^2}$ dB [25]. For $N \in \{2, ..., 10, 12, 15\}$, values of $w_n$ and $a_n$ ($n = 1, ..., N$) are tabulated in [26, Table 25.9]. For $M \in \{2, ..., 10, 12, 16, 20\}$, values of $v_m$ and $b_m$ ($m = 1, ..., M$) are tabulated in [26, Table 25.10]. The Gauss-Hermite series expansion with $M = 12$ is sufficiently accurate [27].

According to (2), the femto DL OP with respect to the SIR target $\gamma_F$ of an indoor FUE at a distance $d$ from the MBS, i.e., $\Pr(SIR_F < \gamma_F | D_{FM} = d)$, is thus lower bounded by

$$\Pr(SIR_F < \gamma_F | D_{FM} = d) \geq F_g\left(\frac{P_{M,Tx} G_M \phi_F r_F^{\alpha_F} \gamma_F}{P_{F,Tx} G_F \phi_{FM} d^{\alpha_{FM}}}; \breve{\mu}_F - \breve{\mu}_{FM}, \sqrt{\tilde{\sigma}_F^2 + \tilde{\sigma}_{FM}^2}\right) + \sum_{n=1}^{N} \sum_{m=1}^{M} \frac{w_n v_m \left\{1 - \exp\left[-\kappa_F \lambda_F \left(e^{\breve{\mu}_F + \sqrt{2a_n + 2\bar{\chi}(b_m)}\tilde{\sigma}_F} - \gamma_F e^{\sqrt{2}\tilde{\sigma}_{FM} b_m + \breve{\mu}_{FM}}\right)^{-\frac{2}{\alpha_{FF}}}\right]\right\}}{2\pi \sqrt{a_n + \bar{\chi}(b_m)} e^{\bar{\chi}(b_m)}} \quad (8)$$

*B. Macrocell Downlink Outage Probability*

For an MUE at a random distance $D_M$ from the MBS, the received SIR of an RB is given by

$$SIR_M = \frac{P_M \phi_M^{-1} D_M^{-\alpha_M} H_M Q_M}{\sum_{i \in \Phi} P_{Fi} \phi_{MF}^{-1} D_{MFi}^{-\alpha_{MF}} H_{MFi} Q_{MFi}} \quad (9)$$

where $\phi_M$ and $\phi_{MF}$ are fixed radio propagation losses, and $\alpha_M$ and $\alpha_{MF}$ are path-loss exponents, as defined in Table I; $D_{MFi}$ ($i \in \Phi$) is the distance from interfering FAP $i$ to the MUE; $H_M$ and $H_{MFi}$ are exponentially distributed unit-mean channel power gains [9] from the MBS and interfering FAP $i$ to the MUE, respectively; $Q_M \sim \text{LN}(\zeta\mu_M, \zeta^2\sigma_M^2)$ and $Q_{MFi} \sim \text{LN}(\zeta\mu_{MF}, \zeta^2\sigma_{MF}^2)$ denote the LN shadowing from the MBS and interfering FAP $i$ to the MUE, respectively, with $\zeta = 0.1\ln10$ [9]. The RVs $H_M$, $Q_M$, $D_M$, $H_{MFi}$, $Q_{MFi}$ and $D_{MFi}$ ($i \in \Phi$) are mutually independent.

As a basic QoS requirement, $\text{SIR}_M$ in (9) needs to be no less than an SIR target $\gamma_M$. Assuming identical channel statistics across all RBs, the DL OP of an outdoor MUE is given by

$$\Pr(\text{SIR}_M < \gamma_M) = \Pr\left(\frac{S_M}{\sum_{i \in \Phi} P_{Fi}\phi_{MF}^{-1}D_{MFi}^{-\alpha_{MF}} H_{MFi}Q_{MFi}} < \gamma_M\right) \quad (10)$$

where $S_M = P_M\phi_M^{-1}D_M^{-\alpha_M} H_M Q_M$ is the received signal power from the MBS. There is no exact closed-form expression of (10) except for the special case of $\alpha_{MF} = 4$ [28]. As in the analysis of the femto DL OP, a lower bound of (10) is given by the probability that at least one dominant interfering FAP is able to individually cause a macro DL outage relative to the SIR target $\gamma_M$, i.e.,

$$\Pr(\text{SIR}_M < \gamma_M) \geq \int_0^\infty \Pr(\Phi_w \neq \varnothing) dF_{S_M}(w) \quad (11)$$

where for any received macro signal power $w$ ($> 0$), $F_{S_M}(w) = \Pr(S_M \leq w)$ is the CDF of $S_M$, $\Phi_w = \{i \in \Phi | w/(P_{Fi}\phi_{MF}^{-1}D_{MFi}^{-\alpha_{MF}} H_{MFi}Q_{MFi}) < \gamma_M\}$ denotes the set of dominant interfering FAPs to the outdoor MUE of interest, and $\Pr(\Phi_w \neq \varnothing)$ is the probability of the set $\Phi_w$ being nonempty.

For an outdoor MUE at a distance $d$ from the MBS, the received macro signal power is given by $S_M(d) = P_M\phi_M^{-1}d^{-\alpha_M}H_M Q_M$. Since $H_M Q_M \sim \text{LN}(\tilde{\mu}_M, \tilde{\sigma}_M^2)$, with $\tilde{\mu}_M = \zeta(\mu_M - 2.5)$ dB and $\tilde{\sigma}_M = \zeta\sqrt{\sigma_M^2 + 5.57^2}$ dB [25], $S_M(d) \sim \text{LN}(\breve{\mu}_M, \tilde{\sigma}_M^2)$, with $\breve{\mu}_M = \tilde{\mu}_M + \ln(P_M\phi_M^{-1}d^{-\alpha_M})$. Substituting the LN PDF of $S_M(d)$ into (11), we obtain the DL OP of an MUE at a distance $d$ from the MBS as

$$\Pr(\text{SIR}_M < \gamma_M | D_M = d) \geq 1 - \sum_{m=1}^{M} \frac{v_m}{\sqrt{\pi}} \exp\left[-\tilde{b}_m \lambda_F \exp\left(\frac{2\zeta\mu_{F,Tx}}{\alpha_{MF}} + \frac{2\zeta^2\sigma_{F,Tx}^2}{\alpha_{MF}^2}\right)\left(\frac{d^{\alpha_M}}{P_{M,Tx}}\right)^{\frac{2}{\alpha_{MF}}}\right] \quad (12)$$

$$\tilde{b}_m = \pi\left(\frac{G_F\phi_M\gamma_M}{G_M\phi_{MF}}\right)^{\frac{2}{\alpha_{MF}}} \exp\left[\frac{2(\tilde{\mu}_{MF} - \tilde{\mu}_M - \sqrt{2}\tilde{\sigma}_M b_m)}{\alpha_{MF}} + \frac{2\tilde{\sigma}_{MF}^2}{\alpha_{MF}^2}\right]$$

for which the full derivation is provided in Appendix B, assuming that each FAP transmits in an RB with a probability of 100%, where $\tilde{\mu}_{MF} = \zeta(\mu_{MF} - 2.5)$ dB and $\tilde{\sigma}_{MF} = \zeta\sqrt{\sigma_{MF}^2 + 5.57^2}$ dB are the mean and standard deviation in the distribution $LN(\tilde{\mu}_{MF}, \tilde{\sigma}_{MF}^2)$ of $H_{MFi}Q_{MFi}$ ($\forall i \in \Phi$) [25].

## IV. INTERPRETATIONS AND FEMTOCELL SELF-REGULATION

In order to avoid service holes, collocated spectrum-sharing macrocell and femtocells have to meet their DL OP constraints over their targeted service areas. For instance, the probability of the instantaneous femto DL SIR less than the SIR target $\gamma_F$ needs to be kept below $\varepsilon_F$ ($0 \leq \varepsilon_F < 1$), i.e., $\Pr(SIR_F < \gamma_F) \leq \varepsilon_F$. Similarly for macro DL SIR, it requires $\Pr(SIR_M < \gamma_M) \leq \varepsilon_M$, with $0 \leq \varepsilon_M < 1$.

### A. $P_{F,Tx,min}$ and $P_{F,Tx,max}$

In Section III-A, the minimum and maximum powers that an FAP transmits in a subcarrier are denoted by $P_{F,Tx,min}$ and $P_{F,Tx,max}$, respectively. The value of $P_{F,Tx,max}$ can be set as $\overline{P}_{F,Tx,max}$, which is the maximum FAP transmit power per subcarrier limited by a mobile network standard.

Note that macro-to-femto interference is typically much more significant than femto-to-femto interference, due to the much higher transmission power of an MBS than that of an FAP and the double-wall partition between neighboring femtocells. This will be verified by simulation results to be presented in Section V. Hence, $P_{F,Tx,min}$ would be mainly determined by macro-to-femto interference. Since $\Pr(S_F/I_{FM} < \gamma_F | D_{FM} = d)$ in (4) monotonically decreases with $P_{F,Tx}$ for given $d$ and $P_{M,Tx}$, $P_{F,Tx,min}$ is chosen as the minimum value of $P_{F,Tx}$ that makes an FUE at macrocell edge meet $\Pr(S_F/I_{FM} < \gamma_F | D_{FM} = r_M) \leq \varepsilon_F$. Solving $\Pr(S_F/I_{FM} < \gamma_F | D_{FM} = r_M) = \varepsilon_F$ for $P_{F,Tx}$, we have

$$P_{F,Tx,min} = \frac{P_{M,Tx} G_M \phi_F r_F^{\alpha_F} \gamma_F}{G_F \phi_{FM} r_M^{\alpha_{FM}} F_\vartheta^{-1}\left(\varepsilon_F; \tilde{\mu}_F - \tilde{\mu}_{FM}, \sqrt{\tilde{\sigma}_F^2 + \tilde{\sigma}_{FM}^2}\right)} \tag{13}$$

where $F_\vartheta^{-1}\left(\varepsilon_F; \tilde{\mu}_F - \tilde{\mu}_{FM}, \sqrt{\tilde{\sigma}_F^2 + \tilde{\sigma}_{FM}^2}\right)$ is the inverse CDF of the LN RV $\vartheta$ evaluated at $\varepsilon_F$.

### B. Minimum Distance of an FAP from an MBS

According to (4), $\Pr(S_F/I_{FM} < \gamma_F | D_{FM} = d)$ is a monotonically decreasing function of $d$ for given

$P_{M,Tx}$ and $P_{F,Tx}$. If the value of $d$ is too small, i.e., if an FUE gets too close to the MBS, it is likely that $\Pr(S_F/I_{FM} < \gamma_F | D_{FM} = d) > \varepsilon_F$, and inevitably $\Pr(SIR_F < \gamma_F | D_{FM} = d) > \varepsilon_F$. Thus, assuming that all FUEs served by an FAP experience identical path loss from an MBS [12], a necessary condition for a femtocell to meet $\Pr(SIR_F < \gamma_F) \leq \varepsilon_F$ is to be at least $d_{FM,min}$ in distance from the MBS, where $d_{FM,min}$ is obtained by solving $\Pr(S_F/I_{FM} < \gamma_F | D_{FM} = d, P_{F,Tx} = \overline{P}_{F,Tx,max}) = \varepsilon_F$ for $d$, i.e.,

$$d_{FM,min} \approx \left[ \frac{P_{M,Tx} G_M \phi_F r_F^{\alpha_F} \gamma_F}{\overline{P}_{F,Tx,max} G_F \phi_{FM} F_g^{-1}\left(\varepsilon_F; \tilde{\mu}_F - \tilde{\mu}_{FM}, \sqrt{\tilde{\sigma}_F^2 + \tilde{\sigma}_{FM}^2}\right)} \right]^{\frac{1}{\alpha_{FM}}} \tag{14}$$

where $\overline{P}_{F,Tx,max}$ is the maximum FAP transmit power in a subcarrier limited by a mobile network standard. UEs within a range less than $d_{FM,min}$ from an MBS should be served by the MBS.

*C. Decentralized Femtocell Self-Regulation*

According to (1), for given $P_{M,Tx}$, $\lambda_F$, $P_{F,Tx,min}$, $P_{F,Tx,max}$ and $D_{FM}$, $SIR_F$ monotonically increases with $P_{F,Tx}$. Hence, there is a lower bound (LB) of the FAP transmit power per subcarrier required for a femtocell at a distance $d$ from the MBS to meet $\Pr(SIR_F < \gamma_F | D_{FM} = d) \leq \varepsilon_F$. This lower bound, namely $P_{F,Tx}^{(LB)}(d)$, is obtained by solving $\Pr(SIR_F < \gamma_F | D_{FM} = d) = \varepsilon_F$ for $P_{F,Tx}$ based on (8), where the non-linear equation in $P_{F,Tx}$ can be readily solved numerically using standard functions such as fsolve in MATLAB® and NSolve in Mathematica®.

Following previous subsections, $P_{F,Tx}^{(LB)}(d)$ would also be mainly determined by macro-to-femto interference and could be approximated by solving $\Pr(S_F/I_{FM} < \gamma_F | D_{FM} = d) = \varepsilon_F$ for $P_{F,Tx}$, i.e.,

$$P_{F,Tx}^{(LB)}(d) \approx \frac{P_{M,Tx} G_M \phi_F r_F^{\alpha_F} \gamma_F}{G_F \phi_{FM} d^{\alpha_{FM}} F_g^{-1}\left(\varepsilon_F; \tilde{\mu}_F - \tilde{\mu}_{FM}, \sqrt{\tilde{\sigma}_F^2 + \tilde{\sigma}_{FM}^2}\right)} \tag{15}$$

Moreover, (15) shows that $P_{F,Tx}^{(LB)}(d)$ is approximately a monotonically decreasing function of $d$ for given $P_{M,Tx}$. Hence, if $d_{FM,min} \leq d \leq r_M$, then $P_{F,Tx}^{(LB)}(r_M) \leq P_{F,Tx}^{(LB)}(d) \leq P_{F,Tx}^{(LB)}(d_{FM,min})$, where $P_{F,Tx}^{(LB)}(r_M) \approx P_{F,Tx,min}$ following (13), and $P_{F,Tx}^{(LB)}(d_{FM,min}) \approx \overline{P}_{F,Tx,max}$ following (14).

According to (12), for given $\lambda_F$ and $P_{M,Tx}$, $\Pr(SIR_M < \gamma_M | D_M = d)$ monotonically increases with $d$, $\mu_{F,Tx}$ and $\sigma_{F,Tx}$, respectively, where $\mu_{F,Tx} \approx [(P_{F,Tx,min})_{dBm} + (P_{F,Tx,max})_{dBm}]/2$ and $\sigma_{F,Tx} \approx$

$[(P_{F,Tx,max})_{dBm} - (P_{F,Tx,min})_{dBm}]/6$ as discussed in Section III-A. If $P_{F,Tx,min}$ is fixed (e.g., in (13)) while $P_{F,Tx,max}$ is considered as a variable, then in order to maintain the successful DL reception of an MUE at a distance $d$ from the MBS, the value of $P_{F,Tx,max}$ needs to be upper bounded (UB) by $P_{F,Tx,max}^{(UB)}(d)$, which is given by solving $\Pr(\text{SIR}_M < \gamma_M | D_M = d) = \varepsilon_M$ for $P_{F,Tx,max}$ based on (12). It is easy to show that $P_{F,Tx,max}^{(UB)}(d) \geq P_{F,Tx,max}^{(UB)}(r_M)$ for $d \leq r_M$, and $P_{F,Tx,max}^{(UB)}(r_M)$ is the upper bound on maximum transmission power per subcarrier of all FAPs in order for the MBS to provide DL services to arbitrarily located MUEs over a disc area of radius $r_M$, for a given spatial density $\lambda_F$.

The above discussions show that there are distance-dependent upper and lower bounds on FAP transmit power for ensuring reliable macro and femto DL services. Considering that macrocells typically have a higher priority to access the available spectrum than closed-access femtocells [13], [29], as macrocells provide the infrastructural network to most mobile UEs, and under the assumption that each FAP knows its distance from the closest MBS [12], we propose the following strategy for an FAP to self-regulate its transmit power and usage of radio resource according to its distance $d$ ($d_{FM,min} \leq d \leq r_M$) from the closest MBS: for a given RB,

a) if $P_{F,Tx}^{(LB)}(d) \leq \min\{P_{F,Tx,max}^{(UB)}(d), \overline{P}_{F,Tx,max}\}$, then the FAP should transmit in the RB with a power per subcarrier properly selected in the range $[P_{F,Tx}^{(LB)}(d), \min\{P_{F,Tx,max}^{(UB)}(d), \overline{P}_{F,Tx,max}\}]$, where $P_{F,Tx,max}^{(UB)}(d)$ is used instead of $P_{F,Tx,max}^{(UB)}(r_M)$, in order to make it possible for FAPs located closer to the MBS to transmit at power levels higher than $P_{F,Tx}^{(LB)}(d)$;

b) if $P_{F,Tx}^{(LB)}(d) > \min\{P_{F,Tx,max}^{(UB)}(d), \overline{P}_{F,Tx,max}\}$, indicating $P_{F,Tx,max}^{(UB)}(d) < P_{F,Tx}^{(LB)}(d) \leq \overline{P}_{F,Tx,max}$, because (15) has shown that $P_{F,Tx}^{(LB)}(d) \leq \overline{P}_{F,Tx,max}$ for $d \geq d_{FM,min}$, then the FAP should only transmit in the RB with a reduced probability $\rho$ ($0 < \rho \leq 1$) at the power level $P_{F,Tx}^{(LB)}(d)$. The probability $\rho$ is designed for maintaining the successful DL reception of an MUE at the macrocell edge by reducing the spatial density of FAPs transmitting in the RB, and is given based on (12) by

$$\rho = \exp\left\{\frac{\varsigma\left[\left(P_{F,Tx,max}^{(UB)}(r_M)\right)_{dBm} - \left(\overline{P}_{F,Tx,max}\right)_{dBm}\right]}{\alpha_{MF}} + \frac{\varsigma^2\left[\left(P_{F,Tx,max}^{(UB)}(r_M)\right)_{dBm}^2 - \left(\overline{P}_{F,Tx,max}\right)_{dBm}^2\right]}{18\alpha_{MF}^2} - \frac{\varsigma^2\left(P_{F,Tx,min}\right)_{dBm}\left[\left(P_{F,Tx,max}^{(UB)}(r_M)\right)_{dBm} - \left(\overline{P}_{F,Tx,max}\right)_{dBm}\right]}{9\alpha_{MF}^2}\right\} \quad (16)$$

where $P_{\text{F,Tx,max}}^{(\text{UB})}(r_{\text{M}})$ is a function of $\lambda_{\text{F}}$, and thus $\rho$ also varies with $\lambda_{\text{F}}$.

By substituting $\rho\lambda_{\text{F}}$ in place of $\lambda_{\text{F}}$ in (12), we can show that $\rho\lambda_{\text{F}}$ is the maximum spatial density of FAPs transmitting in an RB, for maintaining successful DL reception in the RB of an MUE at the macrocell edge with $P_{\text{F,Tx,max}} = \overline{P}_{\text{F,Tx,max}}$, i.e., $\Pr(\text{SIR}_{\text{M}} < \gamma_{\text{M}} | D_{\text{M}} = r_{\text{M}}, P_{\text{F,Tx,max}} = \overline{P}_{\text{F,Tx,max}}) = \varepsilon_{\text{M}}$. The probability $\rho$ can be controlled in a way similar to the F-ALOHA strategy [9].

In this femtocell self-regulation strategy, the computation of $P_{\text{F,Tx}}^{(\text{LB})}(d)$, $P_{\text{F,Tx,max}}^{(\text{UB})}(d)$ and $\rho$ at an FAP requires only infrequent updates of parameters such as the FAP's distance from the closest MBS ($d$), MBS transmit power ($P_{\text{M,Tx}}$), spatial intensity of spectrum-sharing femtocells ($\lambda_{\text{F}}$), and channel statistics. For example, the computation of $P_{\text{F,Tx,max}}^{(\text{UB})}(d)$ based on (12) requires updated information of $d$, $P_{\text{M,Tx}}$, $\lambda_{\text{F}}$, path loss exponents ($\alpha_{\text{M}}$, $\alpha_{\text{MF}}$), fixed propagation losses ($\phi_{\text{M}}$, $\phi_{\text{MF}}$), LN shadowing coefficients ($\tilde{\mu}_{\text{M}}$, $\tilde{\mu}_{\text{MF}}$, $\tilde{\sigma}_{\text{M}}$, $\tilde{\sigma}_{\text{MF}}$), antenna gains ($G_{\text{M}}$, $G_{\text{F}}$), minimum and maximum FAP transmit powers ($P_{\text{F,Tx,min}}$, $\overline{P}_{\text{F,Tx,max}}$, for calculating $\mu_{\text{F,Tx}}$ and $\sigma_{\text{F,Tx}}$), and the macro DL SIR target ($\gamma_{\text{M}}$). Since none of these parameters is likely to change frequently, their values can be provided by the operator through backhaul connection either periodically or in an event-driven manner, e.g., driven by the change of the FAP's location or the MBS transmit power.

## V. SIMULATION AND NUMERICAL RESULTS

In this section, we present simulation and numerical results to evaluate the closed-form lower bounds of femto and macro DL OPs, and the decentralized femtocell self-regulation strategy. Each simulation consists of 100 random drops of indoor closed-access femtocells following a homogeneous SPPP, with an average of $N_{\text{F}}$ (= $\lambda_{\text{F}} \pi r_{\text{M}}^2$) FAPs distributed in the disc area centered at the MBS with a radius $r_{\text{M}}$, and with 1000 trials per drop to simulate random fading and shadowing. Each indoor FUE is $r_{\text{F}}$ away from its serving FAP. Based on 3GPP LTE Release 8 [3], we use a bandwidth of 20 MHz to provide 100 RBs in each DL time slot, with 12 subcarriers per RB. We assume that each FAP knows its distance from the MBS, and the macrocell is fully loaded with all its available RBs in use at any time. DL OPs are evaluated on a per RB basis, but in favor of easy interpretation, simulation and numerical results are presented in terms of total transmit power of an MBS or an FAP. Table II lists the values of major parameters used.

Fig. 1 shows the simulated and analytical DL OPs versus the UE's distance from the MBS, for $N_F = 30, 100$. The FAP of interest transmits at 23 dBm. Transmit power levels per subcarrier of interfering FAPs are generated as independent and identically distributed (i.i.d.) LN RVs, where $P_{F,Tx,min}$ is calculated from (13) and $(\overline{P}_{F,Tx,max})_{dBm} = -7.79$ dBm corresponding to a power of 23 dBm equally distributed among 1200 subcarriers. For a given $N_F$, as the UE's distance from the MBS increases, the macro DL OP increases, while the femto DL OP decreases. At a given distance, the macro DL OP for $N_F = 100$ is much higher than that for $N_F = 30$, but the femto DL OP gets only slightly higher when $N_F$ increases from 30 to 100.

Fig. 2 shows the simulated and analytical DL OPs versus $N_F$, for a UE at 400 m or 800 m from the MBS. The setting of FAP transmit powers is the same as for Fig. 1. At a given distance from the MBS, as $N_F$ increases from 1 to 100, the macro DL OP increases significantly, while the femto DL OP remains almost constant, indicating that femto-to-macro interference is much more significant than femto-to-femto interference. This is mainly due to the double-wall partition between neighboring femtocells that protects indoor FUEs from interfering femto transmissions. This also proves the good approximation of $P_{F,Tx}^{(LB)}(d)$ in (15).

Both Fig. 1 and Fig. 2 show that the femto DL OP lower bound is always in close agreement with simulation results, but the tightness of the macro DL OP lower bound may be affected by a large value of $N_F$ (and equivalently a large $\lambda_F$). This is because the macro DL OP lower bound considers only dominant femto interferers, which may contribute only a part of the total femto-to-macro interference when the number of spectrum-sharing femtocells per cell site is large. It's also worth noting that for reasonably small OP values that we are usually more interested in, e.g., 0.1 and less, the macro DL OP lower bound matches closely with simulation results.

Fig. 3 plots $d_{FM,min}$ calculated from (14), versus the FAP transmit power, for $\xi = 10$ dB, 15 dB.

For either value of $\xi$, $d_{\text{FM,min}}$ decreases with the increase of FAP transmit power. For a given FAP transmit power, $d_{\text{FM,min}}$ is reduced at a higher value of $\xi$, indicating that indoor spectrum-sharing femtocells can be deployed closer to an MBS when the wall-partition loss is higher.

Fig. 4 plots the upper and lower bounds of FAP transmit power required for keeping the macro DL OP relative to $\gamma_M$ below $\varepsilon_M$ and the femto DL OP relative to $\gamma_F$ below $\varepsilon_F$, versus the FAP's distance from the MBS, for $N_F = 30, 100$. The approximate lower bound of FAP transmit power in (15) is also plotted, and is shown to almost overlap the exact lower bound of FAP transmit power. Both the upper and lower bounds of FAP transmit power decrease with the FAP's distance from the MBS. At a given distance, the upper bound of maximum FAP transmit power reduces significantly with the increase of $N_F$, while the lower bound of FAP transmit power does not change much with $N_F$. At distances shorter than 384 m, the required lower bound of FAP transmit power is larger than 23 dBm, which is the maximum transmit power of an FAP. This means that femtocells cannot be deployed at distances less than 384 m from the MBS. For $N_F = 30$, the upper bound is larger than the lower bound at all distances considered, making it possible for each FAP to transmit in an RB with a power per subcarrier in $[P_{F,\text{Tx}}^{(\text{LB})}(d), \min\{P_{F,\text{Tx,max}}^{(\text{UB})}(d),$ $\overline{P}_{F,\text{Tx,max}}\}]$. However, when $N_F = 100$, the upper bound of maximum FAP transmit power falls below the required lower bound of FAP transmission power at distances greater than 384 m, indicating that it is not feasible to overlay 100 spectrum-sharing femtocells on the macrocell.

Fig. 5 shows the simulated DL OPs versus the distance between the UE and the MBS, where for every random drop of femtocells following a homogeneous SPPP, the transmit power level of each FAP was determined following the femtocell self-regulation strategy proposed in Section IV-C, not simply generated as a LN RV. In the simulations, femtocells are deployed only at distances greater than $d_{\text{FM,min}}$ from the MBS. By comparing Fig. 5 with Fig. 1, we can see that the femtocell self-regulation strategy is able to keep the macro DL OP below $\varepsilon_M$ and the femto DL OP not exceeding $\varepsilon_F$ at the same time over the whole targeted service area of the macrocell even for $N_F = 100$. Since the femtocell self-regulation strategy is developed based on our DL OP analysis in Section III, the simulation results in Fig. 5 also verify the feasibility of the i.i.d. LN assumption on transmit power levels of interfering FAPs made in the DL OP analysis.

The associated FAP transmit power and transmission probability $\rho$ in an RB decided by the femtocell self-regulation strategy are plotted against the FAP's distance from the MBS in Fig. 6. We can see that the use of $P_{F,Tx}^{(LB)}(d)$ and $P_{F,Tx,max}^{(UB)}(d)$ in femtocell self-regulation can gradually reduce transmit powers of FAPs that are further away from the MBS, so as to mitigate their interference to nearby MUEs that suffer severe path loss from the MBS. At a given distance from the MBS, when $N_F$ increases from 30 to 100, the femtocell self-regulation strategy not only decreases the FAP transmit power but also reduces the transmission probability $\rho$ from 1 to 0.15. With a total of 100 RBs available per DL time slot, $\rho = 0.15$ means that a femtocell can access 15 RBs at a time, which is still manageable by a femtocell that serves only 2 to 6 UEs [1], [6].

Fig. 7 plots the simulated area spectral efficiency (ASE) (in b/s/Hz/m$^2$) versus $N_F$, when the femtocell self-regulation strategy is employed at each FAP, for $\xi = 10$ dB, 15 dB. The ASE is defined as the network-wide spatially averaged product of the density of successful transmissions subject to an SIR target and the corresponding spectral efficiency [17]. Accordingly, the femto, macro, and overall ASEs are respectively expressed as $ASE_F = E[\rho \lambda_F (1 - OP_F) \log_2(1 + \gamma_F)]$, $ASE_M = E[\lambda_M (1 - OP_M) \log_2(1 + \gamma_M)]$, and $ASE = ASE_F + ASE_M$, where $OP_F = Pr(SIR_F < \gamma_F)$, $OP_M = Pr(SIR_M < \gamma_M)$, $\lambda_M$ is the spatial density of co-channel MUEs on the $\mathbb{R}^2$ plane, and expectations are taken with respect to spatial distributions of UEs and FAPs. Fig. 7 shows that the proposed femtocell self-regulation strategy is able to keep the macro ASE almost unaffected by even a large number of overlaid spectrum-sharing femtocells, and provide an overall ASE much higher than that of the macro network. For $\xi = 15$ dB, the overall ASE and femto ASE increase with $N_F$, indicating that spatial reuse can be improved by deploying more spectrum-sharing indoor femtocells if they are insulated by high wall-partition losses. For $\xi = 10$ dB, the overall ASE and femto ASE start to decrease with $N_F$ when $N_F$ goes beyond a certain value, indicating that if indoor femtocells are not well insulated by surrounding walls, employing too many spectrum-sharing femtocells per cell site may degrade the efficiency of spatial reuse.

## VI. Conclusions

In this paper, we have presented a thorough DL OP analysis for spectrum-sharing macro and femto cells, based on which a decentralized femtocell self-regulation strategy has been proposed for FAPs to adjust their transmit power and usage of OFDMA RBs depending on their locations within the underlying macrocell. Simulation results have shown that the derived closed-form DL OP lower bounds are tight and the femtocell self-regulation strategy is able to ensure satisfactory DL services in targeted macro and femto service areas and provide superior spatial reuse, for even a large number of spectrum-sharing femtocells deployed per cell site. It has been assumed that each FAP knows its distance from the closest MBS, so the presented performance of the decentralized femtocell self-regulation strategy serves as a benchmark. How a femtocell infers its distance from an MBS is beyond the scope of this paper. Due to limited space, we have focused on providing insights on the fundamental limits of DL service provisioning and spatial reuse in collocated spectrum-sharing macro and femto networks. In our future work, we will investigate intelligent user association, resource partitioning, and inter-cell interference coordination schemes that are particularly important in spectrum-sharing macro and femto networks.

## Appendix A

The lower bound of (5) is calculated as follows

$$\begin{aligned}
\Pr\left(\text{SIR}_F < \gamma_F, \frac{S_F}{I_{FM}} \geq \gamma_F\right) &\geq \int_0^\infty \int_0^\infty \Pr(\Phi_{w,z} \neq \varnothing) dF_{S_F}(w) dF_{I_{FM}}(z) \\
&= \int_0^\infty \int_{z\gamma_F}^\infty \left\{1 - \exp\left[-\int_{\mathbb{R}^2} \lambda_F \Pr\left(\frac{w}{z + P_{Fi}\phi_{FF}^{-1}|x|^{-\alpha_{FF}} H_{FFi} Q_{FFi}} < \gamma_F\right) dx\right]\right\} dF_{S_F}(w) dF_{I_{FM}}(z) \\
&= \int_0^\infty \int_{z\gamma_F}^\infty \left\{1 - \exp\left[-2\pi\lambda_F \int_0^\infty \Pr\left(r^{\alpha_{FF}} < \frac{P_{Fi}\phi_{FF}^{-1} H_{FFi} Q_{FFi} \gamma_F}{w - z\gamma_F}\right) r \, dr\right]\right\} dF_{S_F}(w) dF_{I_{FM}}(z) \\
&= \int_0^\infty \int_{z\gamma_F}^\infty \left\{1 - \exp\left[-\pi\lambda_F \left(\frac{G_F G_U \gamma_F}{\phi_{FF}(w - z\gamma_F)}\right)^{\frac{2}{\alpha_{FF}}} \mathbb{E}\left[(P_{Fi,\text{Tx}} H_{FFi} Q_{FFi})^{\frac{2}{\alpha_{FF}}}\right]\right]\right\} dF_{S_F}(w) dF_{I_{FM}}(z) \\
&\approx \int_0^\infty \int_{z\gamma_F}^\infty \left\{1 - \exp\left[-\kappa_F \lambda_F (w - z\gamma_F)^{-\frac{2}{\alpha_{FF}}}\right]\right\} dF_{S_F}(w) dF_{I_{FM}}(z)
\end{aligned} \quad (17)$$

where $\Pr(\Phi_{w,z} \neq \varnothing) = 1 - \exp\left(-\int_{\mathbb{R}^2} \lambda_{w,z}(x) dx\right)$, $\lambda_{w,z}(x) = \lambda_F \Pr\left(w / \left(z + P_{Fi} \phi_{FF}^{-1} |x|^{-\alpha_{FF}} H_{FFi} Q_{FFi}\right) < \gamma_F, w/z \geq \gamma_F\right)$ is the density of points in the set $\Phi_{w,z}$ at location $x$ on the $\mathbb{R}^2$ plane [17]; as $P_{Fi,Tx} H_{FFi} Q_{FFi} \sim \mathrm{LN}(\tilde{\mu}_{FF} + \zeta \mu_{F,Tx}, \tilde{\sigma}_{FF}^2 + \zeta^2 \sigma_{F,Tx}^2)$ [25], we have $\mathrm{E}\left[(P_{Fi,Tx} H_{FFi} Q_{FFi})^{\frac{2}{\alpha_{FF}}}\right] = \exp\left[\frac{2(\tilde{\mu}_{FF} + \zeta \mu_{F,Tx})}{\alpha_{FF}} + \frac{2(\tilde{\sigma}_{FF}^2 + \zeta^2 \sigma_{F,Tx}^2)}{\alpha_{FF}^2}\right]$;

and for simplifying notations, we define $\kappa_F = \pi \left(\frac{G_F G_U \gamma_F}{\phi_{FF}}\right)^{\frac{2}{\alpha_{FF}}} \exp\left[\frac{2(\tilde{\mu}_{FF} + \zeta \mu_{F,Tx})}{\alpha_{FF}} + \frac{2(\tilde{\sigma}_{FF}^2 + \zeta^2 \sigma_{F,Tx}^2)}{\alpha_{FF}^2}\right]$.

Since $S_F \sim \mathrm{LN}(\breve{\mu}_F, \tilde{\sigma}_F^2)$, by substituting the LN PDF of $S_F$ into (17), we have

$$
\begin{aligned}
\Pr\left(\mathrm{SIR}_F < \gamma_F, \frac{S_F}{I_{FM}} \geq \gamma_F\right) &\geq \int_0^\infty \int_{z\gamma_F}^\infty \left\{1 - \exp\left[-\kappa_F \lambda_F (w - z\gamma_F)^{-\frac{2}{\alpha_{FF}}}\right]\right\} \frac{\exp\left[-\frac{(\ln w - \breve{\mu}_F)^2}{2\tilde{\sigma}_F^2}\right]}{\sqrt{2\pi} \tilde{\sigma}_F w} dw \, dF_{I_{FM}}(z) \\
&= \int_0^\infty \int_{\chi(z)}^\infty \left\{1 - \exp\left[-\kappa_F \lambda_F \left(e^{\breve{\mu}_F + \sqrt{2x} \tilde{\sigma}_F} - z\gamma_F\right)^{-\frac{2}{\alpha_{FF}}}\right]\right\} \frac{e^{-x} dx}{2\sqrt{\pi x}} dF_{I_{FM}}(z) \\
&= \int_0^\infty \int_0^\infty \left\{1 - \exp\left[-\kappa_F \lambda_F \left(e^{\breve{\mu}_F + \sqrt{2t + 2\chi(z)} \tilde{\sigma}_F} - z\gamma_F\right)^{-\frac{2}{\alpha_{FF}}}\right]\right\} \frac{e^{-t-\chi(z)} dt \, dF_{I_{FM}}(z)}{2\sqrt{\pi(t + \chi(z))}} \\
&\approx \int_0^\infty \sum_{n=1}^N \frac{w_n \left\{1 - \exp\left[-\kappa_F \lambda_F \left(e^{\breve{\mu}_F + \sqrt{2a_n + 2\chi(z)} \tilde{\sigma}_F} - z\gamma_F\right)^{-\frac{2}{\alpha_{FF}}}\right]\right\}}{2\sqrt{\pi(a_n + \chi(z))} e^{\chi(z)}} dF_{I_{FM}}(z)
\end{aligned}
\tag{18}
$$

where $\chi(z) = (\ln z + \ln \gamma_F - \breve{\mu}_F)^2 / (2\tilde{\sigma}_F^2)$, and the last line is based on the Laguerre integration [26].

Since $I_{FM}(d) \sim \mathrm{LN}(\breve{\mu}_{FM}, \tilde{\sigma}_{FM}^2)$, by substituting the LN PDF of $I_{FM}(d)$ into (18), we have

$$
\begin{aligned}
\Pr\left(\mathrm{SIR}_F < \gamma_F, \frac{S_F}{I_{FM}} \geq \gamma_F \middle| D_{FM} = d\right) &\geq \sum_{n=1}^N \int_{-\infty}^\infty \frac{w_n \left\{1 - \exp\left[-\kappa_F \lambda_F \left(e^{\breve{\mu}_F + \sqrt{2a_n + 2\breve{\chi}(y)} \tilde{\sigma}_F} - \gamma_F e^{\sqrt{2} \tilde{\sigma}_{FM} y + \breve{\mu}_{FM}}\right)^{-\frac{2}{\alpha_{FF}}}\right]\right\}}{2\sqrt{\pi(a_n + \breve{\chi}(y))} e^{\breve{\chi}(y)}} \frac{e^{-y^2} dy}{\sqrt{\pi}} \\
&\approx \sum_{n=1}^N \sum_{m=1}^M \frac{w_n v_m \left\{1 - \exp\left[-\kappa_F \lambda_F \left(e^{\breve{\mu}_F + \sqrt{2a_n + 2\breve{\chi}(b_m)} \tilde{\sigma}_F} - \gamma_F e^{\sqrt{2} \tilde{\sigma}_{FM} b_m + \breve{\mu}_{FM}}\right)^{-\frac{2}{\alpha_{FF}}}\right]\right\}}{2\pi \sqrt{a_n + \breve{\chi}(b_m)} e^{\breve{\chi}(b_m)}}
\end{aligned}
\tag{19}
$$

where $\breve{\chi}(y) = \left(\sqrt{2} \tilde{\sigma}_{FM} y + \breve{\mu}_{FM} + \ln \gamma_F - \breve{\mu}_F\right)^2 / (2\tilde{\sigma}_F^2)$, the last line uses the Gauss-Hermite integration [26].

## APPENDIX B

The lower bound of (10) is calculated as follows

$$\Pr(\text{SIR}_M < \gamma_M) \geq \int_0^\infty \Pr(\Phi_w \neq \varnothing) dF_{S_M}(w)$$

$$= 1 - \int_0^\infty \exp\left[-\int_{\mathbb{R}^2} \lambda_F \Pr\left(\frac{w}{P_{Fi}\phi_{MF}^{-1}|x|^{-\alpha_{MF}} H_{MFi} Q_{MFi}} < \gamma_M\right) dx\right] dF_{S_M}(w)$$

$$= 1 - \int_0^\infty \exp\left[-\pi \lambda_F \left(\frac{G_F G_U \gamma_M}{\phi_{MF} w}\right)^{\frac{2}{\alpha_{MF}}} \mathrm{E}\left[(P_{Fi,\text{Tx}} H_{MFi} Q_{MFi})^{\frac{2}{\alpha_{MF}}}\right]\right] dF_{S_M}(w) \quad (20)$$

$$\approx 1 - \int_0^\infty \exp\left(-\kappa_M \lambda_F w^{-\frac{2}{\alpha_{MF}}}\right) dF_{S_M}(w)$$

where $\Pr(\Phi_w \neq \varnothing) = 1 - \exp\left(-\int_{\mathbb{R}^2} \lambda_w(x) dx\right)$, $\lambda_w(x) = \lambda_F \Pr\left(w < \gamma_M P_{Fi} \phi_{MF}^{-1} |x|^{-\alpha_{MF}} H_{MFi} Q_{MFi}\right)$, $P_{Fi,\text{Tx}} H_{MFi} Q_{MFi} \sim \text{LN}(\tilde{\mu}_{MF} + \zeta \mu_{F,\text{Tx}}, \tilde{\sigma}_{MF}^2 + \zeta^2 \sigma_{F,\text{Tx}}^2)$, and $\kappa_M = \pi\left(\frac{G_F G_U \gamma_M}{\phi_{MF}}\right)^{\frac{2}{\alpha_{MF}}} \exp\left[\frac{2(\tilde{\mu}_{MF} + \zeta \mu_{F,\text{Tx}})}{\alpha_{MF}} + \frac{2(\tilde{\sigma}_{MF}^2 + \zeta^2 \sigma_{F,\text{Tx}}^2)}{\alpha_{MF}^2}\right]$.

Since $S_M(d) \sim \text{LN}(\breve{\mu}_M, \tilde{\sigma}_M^2)$, by substituting the LN PDF of $S_M(d)$ into (20), we have

$$\Pr(\text{SIR}_M < \gamma_M | D_M = d) \geq 1 - \int_0^\infty \exp\left(-\kappa_M \lambda_F w^{-\frac{2}{\alpha_{MF}}}\right) \frac{\exp\left[-\frac{(\ln w - \breve{\mu}_M)^2}{2\tilde{\sigma}_M^2}\right]}{\sqrt{2\pi}\tilde{\sigma}_M w} dw$$

$$= 1 - \int_{-\infty}^\infty \exp\left[-\kappa_M \lambda_F \exp\left(-\frac{2\sqrt{2}\tilde{\sigma}_M z + 2\breve{\mu}_M}{\alpha_{MF}}\right)\right] \frac{\exp(-z^2)}{\sqrt{\pi}} dz \quad (21)$$

$$\approx 1 - \sum_{m=1}^M \frac{v_m}{\sqrt{\pi}} \exp\left[-\kappa_M \lambda_F \exp\left(-\frac{2\sqrt{2}\tilde{\sigma}_M b_m + 2\breve{\mu}_M}{\alpha_{MF}}\right)\right]$$

$$= 1 - \sum_{m=1}^M \frac{v_m}{\sqrt{\pi}} \exp\left[-\tilde{b}_m \lambda_F \exp\left(\frac{2\zeta \mu_{F,\text{Tx}}}{\alpha_{MF}} + \frac{2\zeta^2 \sigma_{F,\text{Tx}}^2}{\alpha_{MF}^2}\right)\left(\frac{d^{\alpha_M}}{P_{M,\text{Tx}}}\right)^{\frac{2}{\alpha_{MF}}}\right]$$

where the Gauss-Hermite integration [26] is used, and $\tilde{b}_m$ is defined in (12).

TABLE I  TERRESTRIAL RADIO PROPAGATION MODEL [19]

| Link (Tx/Rx) | Fixed propagation loss | Path-loss exponent | Wall-partition loss | Path loss |
|---|---|---|---|---|
| MBS/Outdoor UE | $\phi_M = 10^{-7.1}f_c^3$ | $\alpha_M$ | N/A | $PL_M = \phi_M d^{\alpha_M}$ |
| Serving FAP/Indoor UE | $\phi_F = 10^{3.7}$ | $\alpha_F$ | N/A | $PL_F = \phi_F d^{\alpha_F}$ |
| FAP/Outdoor UE | $\phi_{MF} = \phi_F \xi$ | $\alpha_{MF}$ | $\xi$ | $PL_{MF} = \phi_{MF} d^{\alpha_{MF}}$ |
| MBS/Indoor UE | $\phi_{FM} = \phi_M \xi$ | $\alpha_{FM}$ | $\xi$ | $PL_{FM} = \phi_{FM} d^{\alpha_{FM}}$ |
| Interfering FAP/Indoor UE | $\phi_{FF} = \phi_F \xi^2$ | $\alpha_{FF}$ | $\xi^2$ | $PL_{FF} = \phi_{FF} d^{\alpha_{FF}}$ |

In Table I, $f_c$ is the carrier frequency in MHz, $d$ denotes the range of the link, and $\xi^2$ indicates that double wall-partition losses are assumed for all links from interfering FAPs to the indoor UE of interest.

TABLE II SYSTEM AND CHANNEL PARAMETERS

| Parameters | Values | Parameters | Values |
|---|---|---|---|
| Wall-partition loss ($\xi$) | 10 dB (and 15 dB in Fig. 3 and Fig. 7) | MBS transmission power | 43 dBm |
| Outdoor path loss exponent ($\alpha_M$) | 4 | FAP transmission power | $\leq$ 23 dBm |
| Indoor path loss exponent ($\alpha_F$) | 3 | MBS antenna gain ($G_M$) | 15 dBi |
| Indoor-and-outdoor path loss exponents ($\alpha_{FF}$, $\alpha_{MF}$, $\alpha_{FM}$) | 4 | FAP antenna gain ($G_F$) | 2 dBi |
| Outdoor LN shadowing standard deviation ($\sigma_M$) | 8 dB | UE antenna gain ($G_U$) | 0 dBi |
| Indoor LN shadowing standard deviation ($\sigma_F$) | 4 dB | Macrocell radius ($r_M$) | 1000 m |
| Femto-to-femto LN shadowing standard deviation ($\sigma_{FF}$) | 12 dB | Femtocell radius ($r_F$) | 30 m |
| Femto-to-macro LN shadowing standard deviation ($\sigma_{MF}$, $\sigma_{FM}$) | 10 dB | Macro SIR target ($\gamma_M$) | 5 dB |
| LN shadowing means ($\mu_F$, $\mu_{FM}$, $\mu_{FF}$, $\mu_M$, and $\mu_{MF}$) | 0 dB | Femto SIR target ($\gamma_F$) | 10 dB |
| Macro and femto DL OP constraints ($\varepsilon_M$, $\varepsilon_F$) | 0.1 | Carrier frequency ($f_c$) | 2000 MHz |
| Number of MUEs per macrocell | 100 | Number of subcarriers | 1200 |
| Number of FUEs per femtocell | 2 | Number of RBs per DL time slot | 100 |
| Order of the Gauss-Hermite integration ($M$) | 12 | Order of the Laguerre integration ($N$) | 12 |

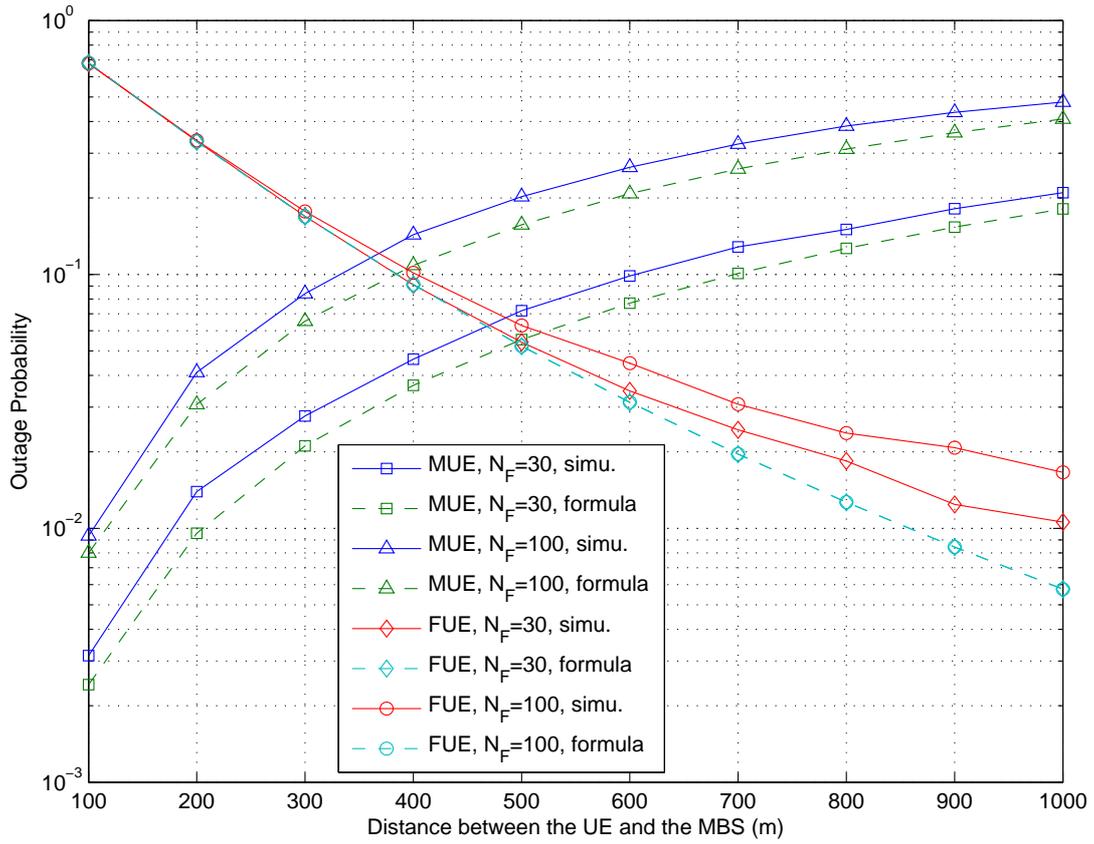

**Fig. 1** Outage probability vs. distance of a UE from the MBS, for $N_F = 30$ and 100, $\xi = 10$ dB, the FAP of interest transmits at a power of 23 dBm, and the maximum transmit power of interfering FAPs is 23 dBm.

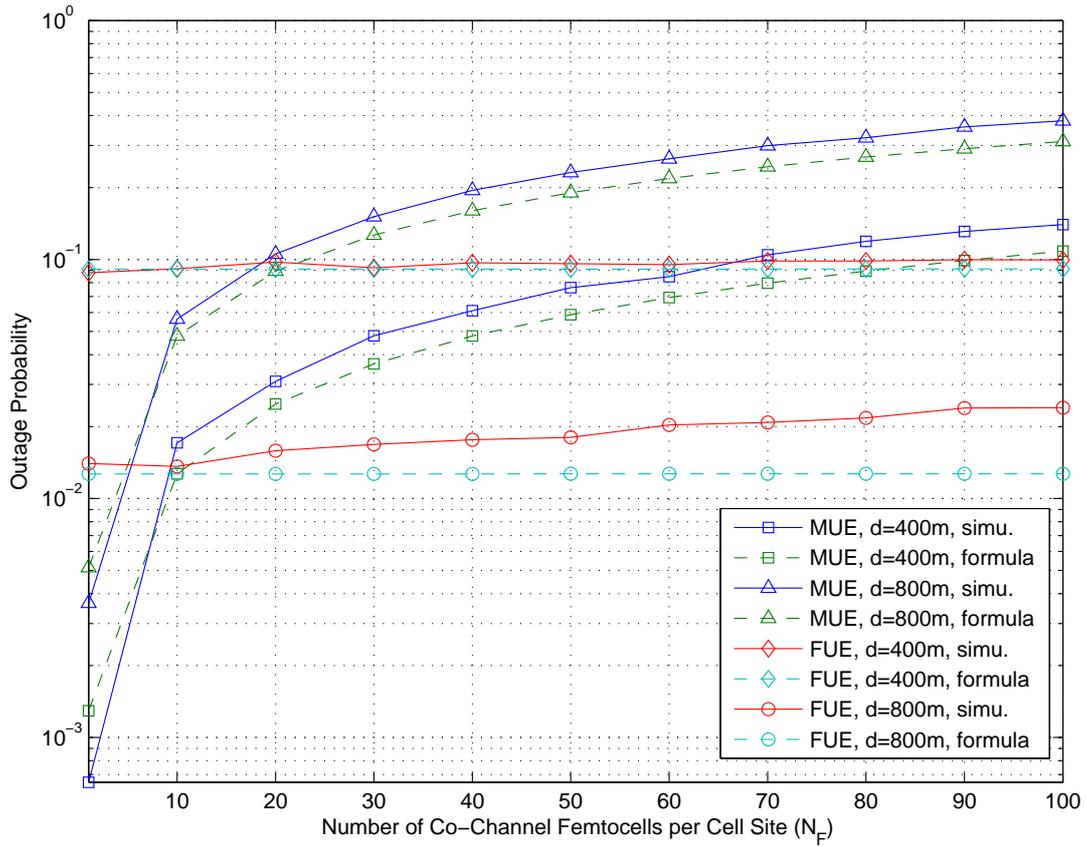

**Fig. 2** Outage probability vs. the number of femtocells per cell site ($N_F$), for a UE at a distance of 400 m or 800 m from the MBS, $\xi = 10$ dB, the FAP of interest transmits with a power of 23 dBm, and the maximum transmit power of all interfering FAPs is 23 dBm.

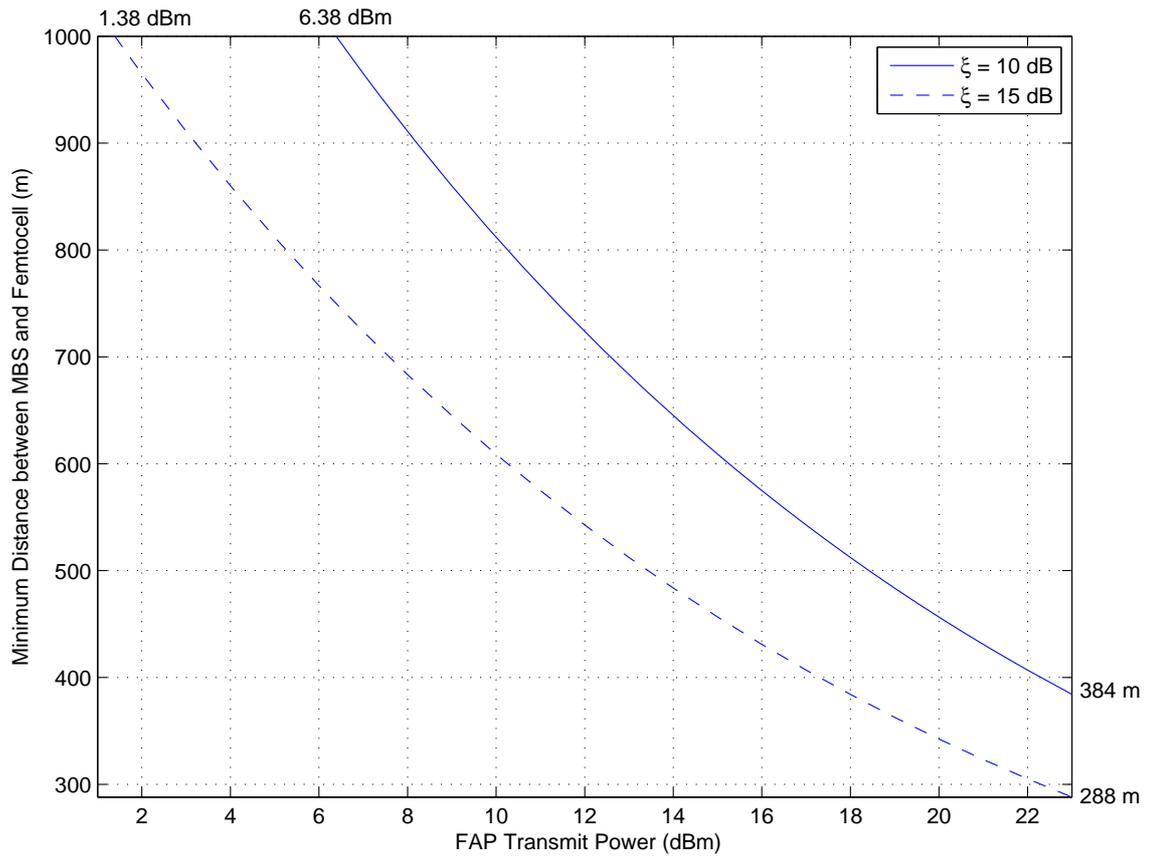

**Fig. 3** $d_{\text{FM,min}}$ vs. FAP transmit power, for $\xi$ = 10 dB and 15 dB.

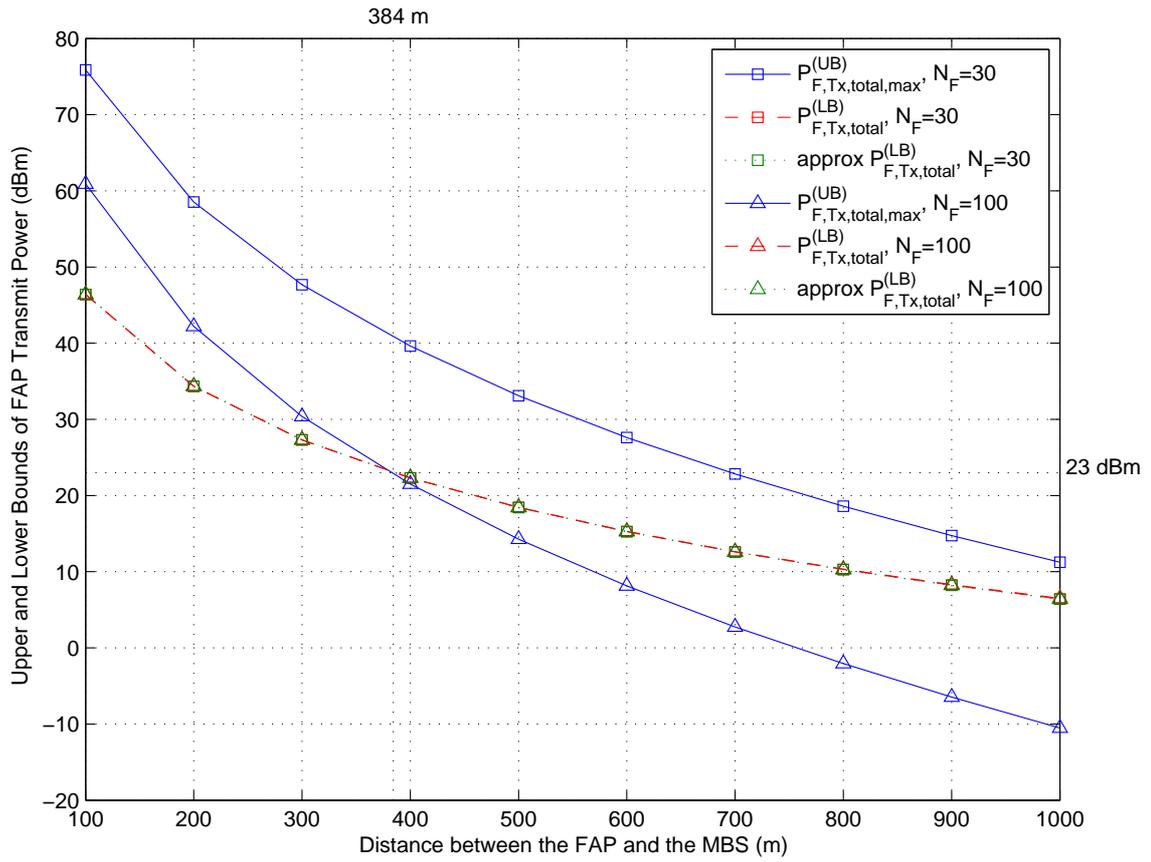

**Fig. 4** Upper bound of maximum FAP transmit power, lower bound of FAP transmit power, and approximate lower bound of FAP transmit power in (15) vs. distance of an FAP from the MBS, for $N_F$ = 30 and 100, and $\xi$ = 10 dB.

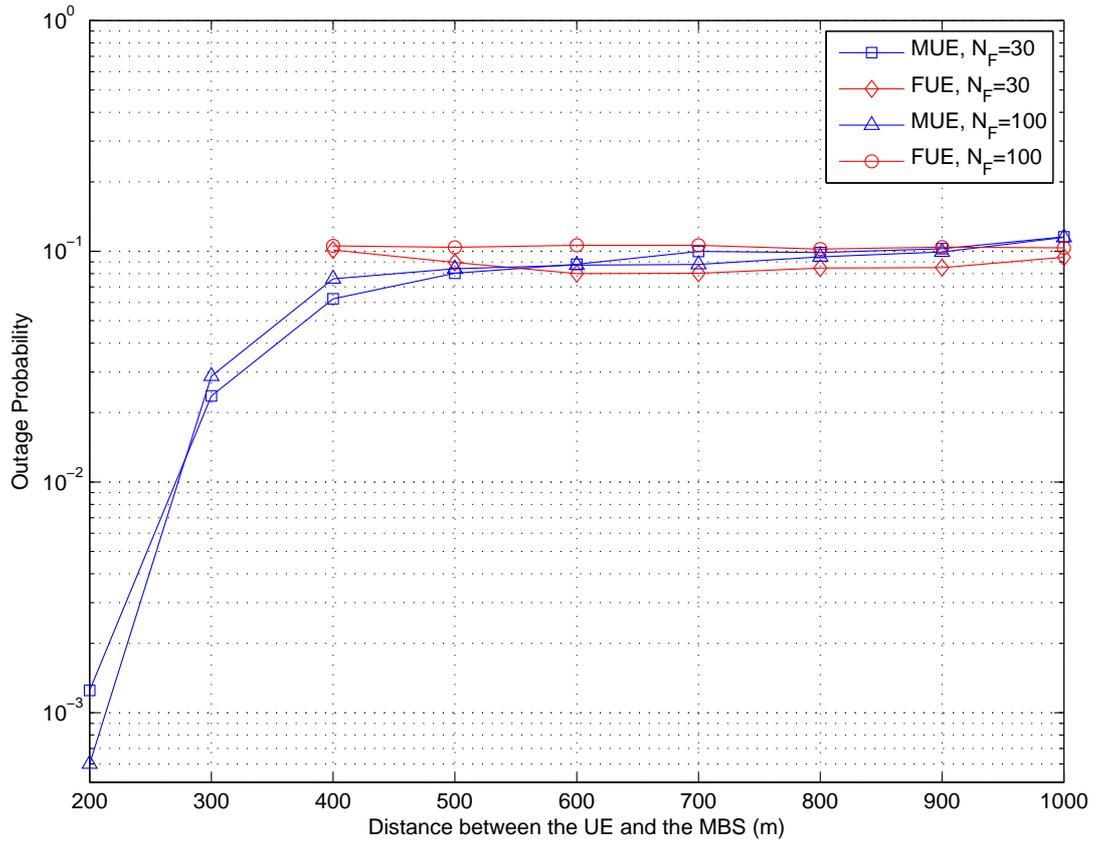

**Fig. 5** Outage probability obtained from simulations of the decentralized femtocell self-regulation strategy, vs. distance of a UE from the MBS, for $N_F$ = 30 and 100, and $\xi$ = 10 dB.

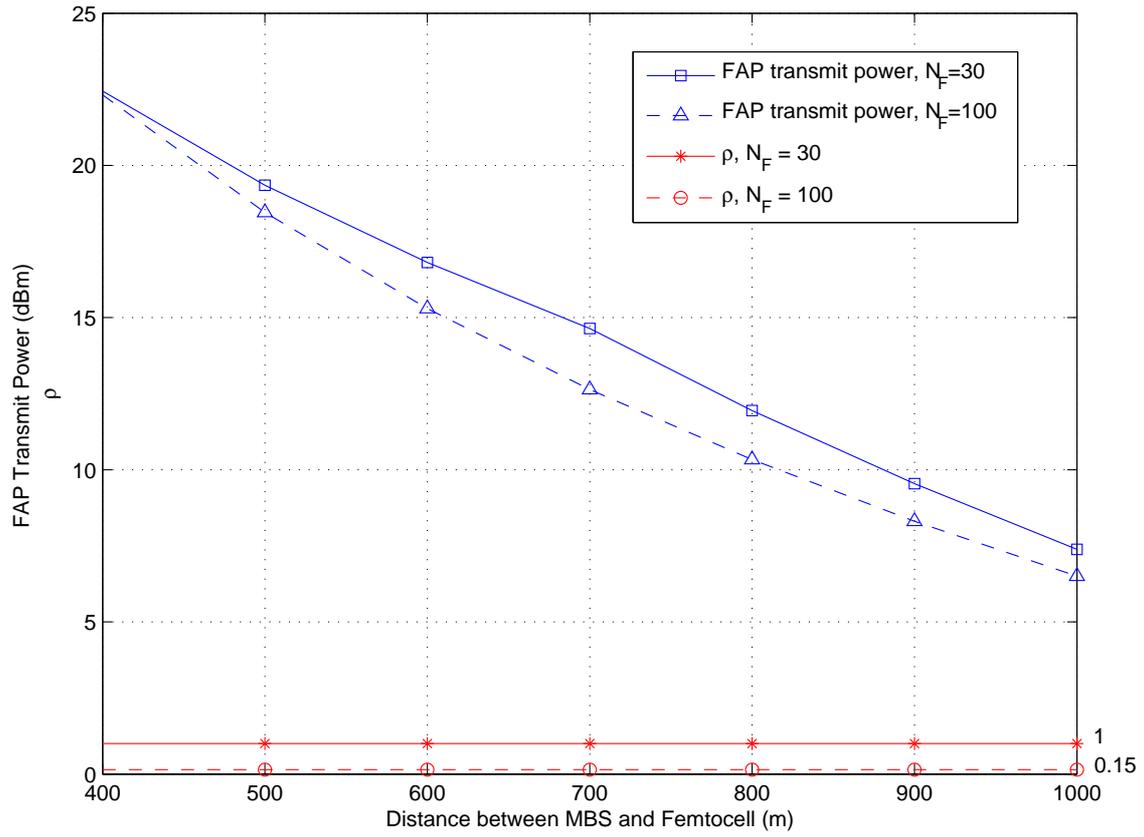

**Fig. 6** FAP transmit power and transmission probability in an RB ($\rho$) given by the decentralized femtocell self-regulation strategy, vs. distance of the FAP from the MBS, for $N_F$ = 30, 100, and $\xi$ = 10 dB.

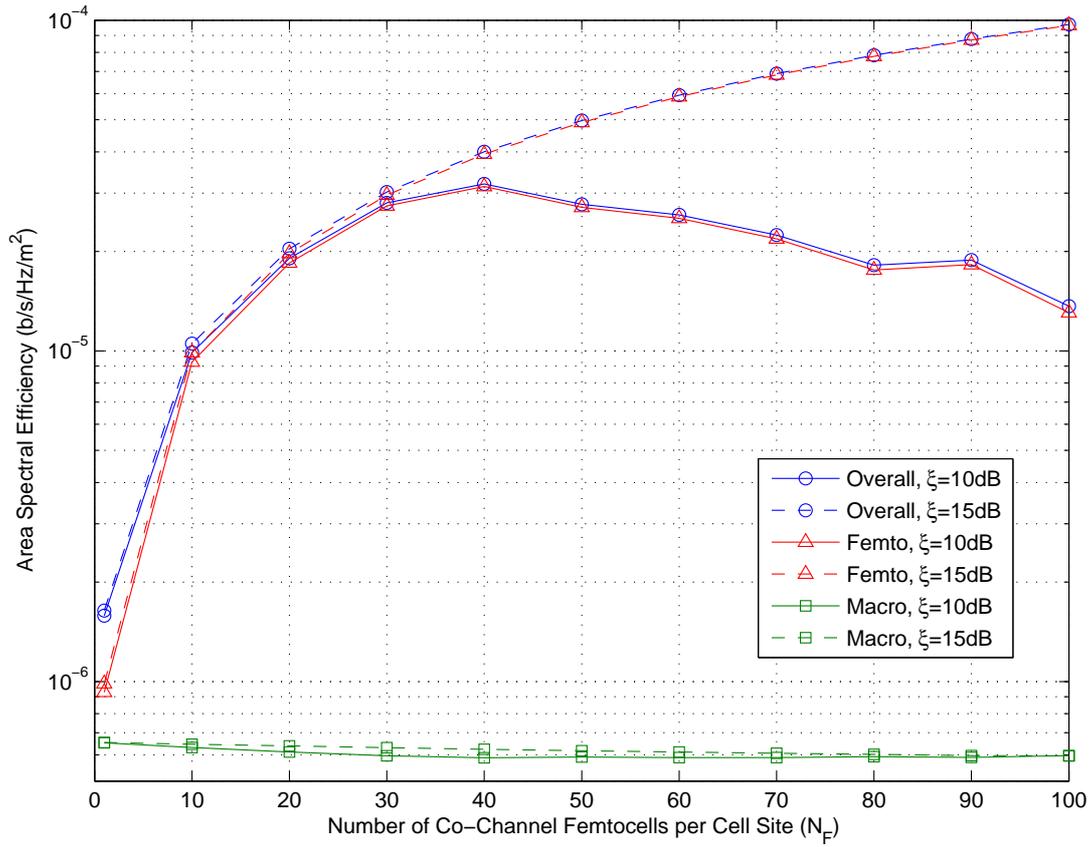

**Fig. 7** ASE obtained from simulations of the proposed decentralized femtocell self-regulation strategy, vs. the number of femtocells per cell site ($N_F$), for $\xi = 10$ dB and 15 dB.